# Participatory Ecosystem Management Planning at Tuzla Lake (Turkey) Using Fuzzy Cognitive Mapping

Filiz Dadaşer Çelik [1], Uygar Özesmi*[1], Asuman Akdoğan [2]

[1] Environmental Science Chair, Department of Environmental Engineering, Faculty of Engineering, Erciyes University, 38039 Kayseri Turkey

[2] Magament and Organization Chair, Department of Business Administration, Faculty of Economics and Administrative Sciences, Erciyes University, 38039 Kayseri Turkey

* Corresponding Author. Tel: 90 352 437 6748 Fax: 90 352 437 4404

email: uozesmi@erciyes.edu.tr

**Abstract**

A participatory environmental management plan was prepared for Tuzla Lake, Turkey. Fuzzy cognitive mapping approach was used to obtain stakeholder views and desires. Cognitive maps were prepared with 44 stakeholders (villagers, local decisionmakers, government and non-government organization (NGO) officials). Graph theory indices, statistical methods and "What-if" simulations were used in the analysis. The most mentioned variables were livelihood, agriculture and animal husbandry. The most central variable was agriculture for local people (villagers and local decisionmakers) and education for NGO&Government officials. All the stakeholders agreed that livelihood was increased by agriculture and animal husbandry while hunting decreased birds and wildlife. Although local people focused on their livelihoods, NGO&Government officials focused on conservation of Tuzla Lake and education of local people. Stakeholders indicated that the conservation status of Tuzla Lake should be strengthened to conserve the ecosystem and biodiversity, which may be negatively impacted by agriculture and irrigation. Stakeholders mentioned salt extraction, ecotourism, and carpet weaving as alternative economic activities. Cognitive mapping provided an effective tool for the inclusion of the stakeholders' views and ensured initial participation in environmental planning and policy making.

# 1. Introduction

Customary approaches to ecosystem conservation were more centralized and decision-making for environmental management was under the responsibility of government officials and technical experts (Glicken, 1999). Customary approaches were based on the assumptions that "local peoples' stakes and rights in environmental issues were subsidiary of those of state" (Borrini-Feyerabend, 1997) and local people did not have enough technical knowledge to contribute to decision-making. Local people and local knowledge were often overlooked by the environmental planners and were not considered during decision-making process. Local knowledge, however, may provide valuable information because people living around an ecosystem have substantial amount of local knowledge, culturally transmitted and accumulated over generations (Berkes and Folke, 1998). Although participation was an integral part in traditional natural resource management systems, these systems were ignored in modernist state projects (Scott, 1998).

In Turkey, the situation was the same until a participatory management plan for Uluabat Lake was developed in 2003 (Özesmi and Özesmi, 2003; Özesmi, 2003). A similar approach is now being employed by the Biodiversity and Natural Resource Management Project supported by the Global Environmental Facility (GEF) Large Grant in Turkey in four pilot areas, Sultan Marshes, neli Ada, Köprülü Canyon and Cami li protected areas. Therefore, it is important to formalize and apply methodologies that inform and contribute to participatory ecosystem conservation in protected areas in Turkey and elsewhere.

In recent years, environmental management has been evolving more rapidly to consider human beings as components of the ecosystem and participatory environmental management approaches have become more popular in ecosystem conservation. Here ecosystem is defined as an area of The Biosphere defined according to some purpose that includes the interacting components of air, land, water, and living organisms, including people (Vallentyne and Beeton, 1988). By acknowledging that people are part of the ecosystem, the socio-economic and cultural aspects of the ecosystem are considered as well as ecological principles (Borrini-Feyerabend, 1997). This approach seems to be successful in environmental management since recent practices proved that environmental conservation excluding local people could not be successful in the long term and local people often resisted top-down conservation policies (Özesmi, 1999a, 1999b).

Participatory environmental management can be defined as the inclusion of all stakeholders in environmental planning and decision-making. In this definition, stakeholders are the individuals or groups who can affect the achievement or are affected by the achievement of a conservation project's objectives (De Lopez, 2001). In other words stakeholders may include local people who have a close relationship with the ecosystem such as people who farm, graze animals, or extract natural products for food and other uses, local decisionmakers, non-government organization representatives interested in environmental conservation, and government officials who are responsible for environmental policy-making.

Participation of stakeholders can be achieved by the inclusion of their views and desires in the environmental management planning process. In this scheme, participation means the identification of perceptions of different stakeholders including local people, non-government organization (NGO) representatives and government officials and analyzing and synthesizing their ideas to arrive at an environmental management plan.

The shift from customary management approaches to participatory ones in environmental management planning in Turkey has resulted from past conservation projects that have failed. Although stakeholders` participation is mentioned in a number of recent government regulations (e.g. Wetland Conservation Regulation of 2002) as a legal requirement, there were a limited number of studies carried out that included stakeholder views (such as Özesmi, 1999a, 1999b, 2001a, 2001b; Özesmi and Özesmi, 2003; Dada er and Özesmi, 2002). In contrast, participatory management approaches have found implications in several fields in environmental management and there is considerable amount of literature on the use of participatory methods in forest management (Ribot, 1995), fisheries management (Hughey et al., 2000), coastal management (Makoloweka and Shurcliff, 1997), lake, wetland and watershed management (Korfmacher, 2000), environmental impact assessment (Richardson et al., 1998; Palern, 1999), urban environment management (Ogu, 2000) and waste management (Kuniyal et al., 1998) to list a few. In addition, evaluation of participatory programs (Buchy and Race, 2001; Johnson et al., 2001), identification of factors affecting people's participation (Lise, 2000; Webler et al., 2003) and determination of conditions for success and pitfalls in participatory processes (Glicken, 1999; Glicken, 2000, Korfmacher, 2001) have found remarkable place in the literature.

Participatory practices are proven to be successful since they provide higher quality decisions and greater commitment by stakeholders to these decisions (Sample, 1993). They ensure representation of diversity of the community, incorporate local knowledge, experience and creativity and clarify and stabilize communication between stakeholder groups (Kapoor, 2001). Decision makers become more competent through the generation of better decisions with more available information (Glicken, 1999). Glicken defines three types of information that can be obtained through participatory processes. Cognitive knowledge is based on technical expertise and is the type of knowledge presented by scientists and other experts. Experiential knowledge is knowledge based on common sense and personal experience and is usually contributed by local people residing in and using the resources of the ecosystem. Value-based knowledge, also known as social or political knowledge, is moral or normative and relates to how people view the 'goodness' of activities. All these types of information can be compiled using a formalized participatory methodology in planning and policy-making.

Participatory processes allow contributions from local people who have information on the ecosystem as a result of their traditional living styles (Özesmi, 1999a). Local people's knowledge is considered besides technical knowledge of scientists (Finlayson and McCay, 1998), which leads to consideration of the socio-economic and cultural aspects of the ecosystem besides ecological aspects.

At the same time participatory processes are not without pitfalls. Failures in participatory practices can occur for several reasons including a lack of planning and foresight by project managers, misunderstanding and distrust among the stakeholders because technical people and the public speak different languages, participants becoming disillusioned with the process because their input is not taken seriously, lack of communication between project managers and all stakeholders throughout the whole process, and exclusion of key stakeholders from the process (Glicken, 2000). Extent of power-sharing is one of the important determinants of the success because in most cases quality of the program is dependent on the power relations between stakeholders (Kapoor, 2001).

Nevertheless if a formalized participatory methodology is used the process can be successful. The planning approach for participatory management generally contains six steps, which will be followed in this study to some extent. These are determination of stakeholder groups, obtaining stakeholder views and desires, evaluation of the data obtained, preparation of the draft management plan, presentation of the plan to the stakeholders and doing necessary revisions according to their reactions.

The purpose of this research project was to use a participatory approach to develop a draft environmental management plan for Tuzla Lake ecosystem, one of Turkey's important wetlands which is under several threats. This draft plan will then be used as a basis for further participatory processes to arrive at a strategic ecosystem management plan. In this paper we show the utility of fuzzy cognitive mapping in obtaining stakeholder views of and desires for Tuzla Lake ecosystem, and in comparing the views of different stakeholders. By obtaining the views of the stakeholders we were better able to understand both the barriers and opportunities available for conservation of Tuzla Lake. Such a fuzzy cognitive mapping approach may be applicable for other ecosystems where it is desired to create an ecosystem management plan with stakeholder participation and input.

## 2. Tuzla Lake ecosystem

The study area, Tuzla (Palas) Lake, is a relatively pristine saline playa lake located at the bottom of Palas Plain in Central Anatolia in Turkey (39° 02' N, 35° 49' E, 1120 m above sea level). Around the south and southeast parts of the lake there are fresh water and salt-water wetlands and wet meadows showing high habitat diversity. There is a created wetland called Yerta ın Marshes southeast of Tuzla Lake, which was formed by the villagers for cattle and especially water buffalo. Yerta ın Marshes is also an important habitat for breeding birds. For this project we define the Tuzla Lake ecosystem as the drainage basin of Tuzla Lake together with the interacting components of air, land, water, and living organisms, including people. The lake is a natural conservation area declared by the state and under conservation through the Wetland Conservation Regulation of 2002 for its high biodiversity value. The regulation mandates that a participatory management plan to be developed for the conservation of the wetland ecosystem. This study is the initial step for helping that mandate to be realized in a participatory way.

## 3. Methods

We compiled all existing literature on the lake (Magnin and Yarar, 1997; Somuncu, 1996; Schekkerman and van Roomen, 1993). Interviews were conducted with the stakeholders using a fuzzy cognitive mapping approach explained by Özesmi and Özesmi (2004). Stakeholders were determined to include all the groups related to Tuzla Lake. The cognitive maps obtained from stakeholders were coded and analysed using graph theory indices, statistical methods, and "What-if" simulations. Based on the analysis a draft management plan was prepared.

*3.1. Why use fuzzy cognitive maps?*

Fuzzy cognitive mapping is an approach used to determine the perceptions and understandings of different people and stakeholder groups. The term cognitive map refers to a causal model made of variables and connections and is often mistaken with geographical representations of places. In this sense cognitive maps represent local knowledge systems as told by the informants (Özesmi, 1999a) and provide informants' cognitive models about the system. The main assumption of this approach is that individuals have cognitive models that are internal representations of a partially observed world (Bauer, 1975).

Fuzzy cognitive mapping can offer various advantages over other participatory research methods such as questionnaire surveys, structured and unstructured interviews, and mapping and modeling methods such as land-use mapping, resource mapping or historical mapping (Özesmi and Özesmi, 2004). First, fuzzy cognitive mapping provides easier quantitative representation and a number of quantitive tools such as neural network simulations to analyze and prioritize the concepts developed. Second, with fuzzy cognitive mapping the important concepts and relationships are drawn on the map by the interviewees themselves, thus removing potential researcher bias as well as reducing the amount of time spent on analysis after the interviews. Fuzzy cognitive maps are different than other mapping exercises in that it focuses on cognitive models of people rather than spatial information and relationships.

Fuzzy cognitive mapping has not been widely used in environmental planning and decision-making. Radomski and Goeman (1996) used fuzzy cognitive mapping to develop alternatives to improve decision-making in sports-fisheries management. Fuzzy cognitive mapping was used to compare views of different stakeholder groups (Özesmi, 1999a, 199b; Dada er and Özesmi, 2002) and in developing participatory management plan (Özesmi and Özesmi, 2003). Hobbs et al. (2002) used fuzzy cognitive mapping as a tool for defining management objectives for the Lake Erie ecosystem. Mendoza and Prahbu (2003) analyzed linkages and interactions between indicators, obtained from multi-criteria approach to sustainable forest management, using cognitive mapping. Hjortsø (2004) demonstrated the use of fuzzy cognitive mapping to increase stakeholder participation in forest management. Özesmi and Özesmi (2004) formalized the methodology for a multi-step fuzzy cognitive mapping approach for natural resource management.

*3.2. Obtaining cognitive maps of stakeholders*

The interviews started with the introduction of the method to the informant with an out of context sample map. After the informant understood the process, we asked an open-ended question. This question was "What are the variables and parameters related to Tuzla Lake and the people living around? How do these variables affect each other?" After the informant listed the variables, the variables were drawn on a paper and circled. The informant was asked to show the causal connections between these variables. Informants showed the direction of causal connections with arrows and +/- signs and defined the strength of the relationship as "low, medium and high". These statements were than transferred to numerical values as " +/- 0.25, 0.5, and 1".

A total of 44 interviews were conducted that lasted 20-165 minutes. Of the 44 maps prepared, 10 were drawn by local decision-makers, 24 were drawn by villagers, 5 were drawn by government officers and 5 were drawn by non-government organization representatives (Table 1). The villages and municipalities chosen to be included were those where inhabitants had direct impacts on Tuzla Lake. For example, inhabitants of the selected villages, have agricultural land or graze their animals around the lake, or extract salt. First the village heads, municipality heads and the heads of the municipality neighborhoods were interviewed. They helped the researchers to reach other people who could talk about Tuzla Lake. The number of people interviewed from each village was related to the population of that village; more people were interviewed in larger villages. The NGO and government officials interviewed were the people who had projects on Tuzla Lake and thus were familiar with the ecosystem.

The sufficiency of number of interviews was determined by drawing an accumulation curve. To draw this curve a presence/absence matrix of the all variables versus the interviews was used and the order of interviews was randomly selected 200 times in Estimates (a freeware software package which computes randomized accumulation curves) (Colwell, 1997). The accumulation curve was prepared by plotting the graph of new variables against the number of interviews. New variables mentioned went below 2 after the 30th interview and stabilized at about 1 new variable thereafter (Figure 1). Therefore we concluded we had sampled variables sufficiently.

*3.3. Analyzing cognitive maps*

The cognitive maps were transformed into square adjacency matrices. Graph theory indices (density, indegree, outdegree, complexity, centrality, hierarchy index) were calculated using these matrices. The density index shows how connected or sparse the maps are. Density is equal to the number of connections divided by the maximum number of connections possible between these variables (Hage and Harary, 1983). Indegree and outdegree indices can be used to determine whether a variable is a transmitter, receiver variables or an ordinary variable (Harary et al., 1965; Bougon et al., 1977; Eden et al., 1992). Outdegree is the cumulative strength of the connections exiting the variable and indegree equals to the cumulative strength of the connections entering the variable. When outdegree is positive and indegree is zero, the variable is a transmitter variable. When outdegree is zero and indegree is positive, the variable is a receiver variable. If both of them are positive, the variable is an ordinary variable. A large number of receiver variables show the outcomes and implications of the cognitive maps (Eden et al, 1992). Whereas, large number of transmitter variables indicate a "formal-hierarchical" system (Simon, 1996). The complexity of a cognitive map is the ratio of the number of receiver variables to the number of transmitter variables. Larger ratios indicate more complex maps. Centrality (indegree + outdegree) shows the contribution of a variable in a cognitive map. The hierarchy index (MacDonald, 1983) shows whether a cognitive map is democratic or hierarchical. The map is fully hierarchical when the hierarchy index is 1 and is fully democratic when it is zero. Detailed information on coding maps and calculating graph theory indices can be found in Özesmi and Özesmi (2004, pp. 49-52).

We used standard statistical methods to compare graph theory indices of different stakeholder groups. Normality of samples was determined by Kolmogorov-Smirnov and Shapiro-Wilk tests of normality. When the samples were normal, t-test was used and when the samples were not normal, Mann-Whitney test was used. Similarities between cognitive maps were determined by calculating Phi and Yule Q coefficients.

Social cognitive maps of stakeholder groups were prepared by augmenting individual cognitive maps and adding them together. Social cognitive maps are more than a simple sum of all separate maps (Laszlo et al., 1996). Connections having different signs cancel each other and the connections having the same signs become stronger (Özesmi, 1999a). Social cognitive maps prepared for different stakeholder groups allow us to examine the similarities and differences in the perceptions of each group.

Creating condensed or simplified cognitive-maps, provides a better understanding because cognitive maps are complex systems having many variables and connections (Özesmi, 1999a). In condensed maps, related variables were gathered in the same group (Harary et al, 1965; Simon, 1996).

As a final analysis fuzzy cognitive map simulations (Kosko, 1987) were conducted to determine the ecosystem's steady state according to the views of the stakeholders. In addition, "what-if" scenarios were run to determine the trajectory of the ecosystem based on the ecosystem model all the stakeholders, the social cognitive map. A vector of initial states of variables was multiplied with the adjacency matrix of the cognitive map (Kosko, 1987). The results were transformed to the interval [0, 1] using a logistic function, which was $1/(1+e^{-10x})$. The transformation provides a better understanding and representation of activation levels of variables and enables us to compare qualitatively the causal output of variables. To run "what-if" scenarios, specific variables related to the scenario were set at a desired value (0 or 1) at each simulation step (Kosko, 1987). The increase or decrease in the variable value relative to the steady state was then determined (see Özesmi and Özesmi, 2004; p.54-55 for detailed method and calculations).

**4. Results and discussion**

*4.1. Graph theory indices*

In the 44 maps analyzed, the average number of variables is 24.9 ± 8.4 SD in the range of 10-47 variables. The average number of transmitter variables is 9.4 ± 3.8 SD, average number of receiver variables is 7.8 ± 3.1 SD and average number of ordinary variables is 7.8 ±5.1 SD per map. The maps have on average 31.2 ± 26.6 SD connections that result in a density of 0.051 ± 0.017 SD. The hierarchy indices are on average 0.036 ± 0.036 SD.

The comparison of cognitive maps of the stakeholder groups shows that the Phi values are the largest between local decision-makers and villagers and between NGO and government officials (Table 2). Yule Q coefficient values also support these two clusters, indicating that the most similar groups are local decision-makers and villagers, and NGO and government officials. Phi and Yule Q coefficient results prompted us to decide *a priori* to pool local decision-makers and villagers and form a group as local people and to pool NGO and government officials. Phi and Yule Q values were calculated for the new groups. Results show that Phi and Yule Q are smallest between local people and NGO&Government officials indicating these groups are most dissimilar (Table 2). In the Kizilirmak Delta and Uluabat Lake, NGO and government officials were also the most similar groups and that pooled NGO&Government officials and villagers were the most dissimilar groups (Özesmi, 1999a, 1999b; Özesmi and Özesmi, 2003).

The graph theory indices calculated for each stakeholder group is given in Table 3. Since the most dissimilar groups are local people and NGO&Government officials, we compared only their indices statistically. Results show that number of connections and connection/variable ratio are significantly higher in NGO&Government officers than local people (Mann-Whitney test, p=0.037 for number of connections and p=0.014 for connection/variable ratio comparisons), which indicate that NGO&Government officials have a more complex view of the system and defined more connections between variables. In the Kizilirmak Delta, NGO& Government officials also had a significantly higher connection/variable ratio than local people (Özesmi, 1999a, 1999b). However, in the Kizilirmak Delta the local people had significantly more variables in their maps while the number of connections was similar between the two groups. The same general relations between stakeholders also hold true for the Uluabat Lake (Özesmi and Özesmi, 2003).

*4.2. Most mentioned variables*

The variables which are most mentioned in the cognitive maps indicate the variables which are shared by stakeholders (Özesmi and Özesmi, 2004). The most mentioned variables in the social cognitive map of all stakeholders are livelihood, agriculture, animal husbandry and salt extraction (Table 4). The most mentioned variables for local people and NGO&Government officials are the same. Hunting is also among the ten most used variables for each group (Table 4).

These results show the relationship of economic issues with the ecosystem. All the stakeholders of Tuzla Lake ecosystem are focused on livelihood similar to the Kızılırmak Delta, Uluabat Lake and Sultan Marshes ecosystems stakeholders (Özesmi, 1999a, 1999b; Dadaşer and Özesmi, 2002; Özesmi and Özesmi, 2003). Other wetland ecosystems of Mediterranean countries are also very important economically for the local people (Benessaiah, 1988).

*4.3. Most central variables*

The most important variables in the cognitive maps can be determined by looking at centrality values (Özesmi and Özesmi, 2004). Table 5 shows the most central variables in the social cognitive map of all stakeholders. Centrality values indicate that the most important variable is agriculture. Agriculture is affected more by the other variables than its effect on them (indegree>outdegree). The second and the third most important variables are livelihood and animal husbandry, which have the same characteristics as agriculture. Other central variables are Tuzla Lake, drought, education and salt extraction.

When the ten most central variables of the stakeholder groups are compared (Table 6), the results show that they all mention livelihood, agriculture and animal husbandry. Although local people mention salt extraction, drought, salt and some negative impacts of Tuzla Lake such as fog formation, air pollution due to winds raising dust and water level rise, NGO&Government officials focused on education, conservation issues, ecosystem balance, biodiversity and tourism (Table 6).

*4.4. Condensed social cognitive maps*

To simplify cognitive maps condensed social cognitive maps of stakeholder groups were prepared. In condensation, the total number of 206 variables were gathered into 18 condensed variables which were ecosystem integrity, livelihood, agriculture, animal husbandry, salt extraction, carpet weaving, mud, tourism, birds and wildlife, hunting, water problems, water projects, government support, education and socio-cultural structure, conservation of Tuzla Lake, drainage of Tuzla Lake and negative impacts of Tuzla Lake.

The condensed social cognitive map of all stakeholders, reveal that agriculture increases livelihood strongly (Figure 2). Other important connections are animal husbandry and drainage of Tuzla Lake increasing livelihood; negative impacts of the Tuzla Lake increasing ecosystem integrity and ecosystem integrity increasing birds and wildlife and tourism. According to the map with all stakeholders, conservation of Tuzla Lake increases tourism. The strongest negative connections are water problems decreasing agriculture, animal husbandry and salt extraction. Hunting decreases birds and wildlife and agriculture decreases conservation of Tuzla Lake.

*4.5. Similarities and differences between condensed social cognitive maps of stakeholder groups*

Both the local peoples' and NGO&Government officials' condensed social cognitive maps indicate that agriculture and animal husbandry strongly increase livelihood while water problems strongly decrease agriculture and animal husbandry. Although local people mention that salt extraction increases animal husbandry, NGO&Government officials do not draw this connection strongly. According to both groups hunting decrease birds and wildlife while education increases agriculture.

Stakeholder groups have different views on the conservation of Tuzla Lake. Local people think that drainage of Tuzla Lake increases livelihood while agriculture increases drainage of Tuzla Lake, whereas NGO&Government officials indicate that agriculture decreases ecosystem integrity and birds and wildlife. According to NGO&Government officials, conservation of Tuzla Lake increases tourism and decreases hunting. NGO&Government officials agree that education of local people decreases hunting and it increases conservation of Tuzla Lake.

Filho (1997) notes the benefits of integrating environmental education into conservation projects and participatory environmental management. In particular he focuses on the necessity of training environmental managers in environmental education so that they may effectively integrate education into participatory management projects. Stakeholders in Australia and the USA often cited education as one of the benefits of participatory projects (Margerum, 1999). Education activities provide a mutually acceptable goal, publicity for the project, and an immediate action. Based on the results of our research, a local NGO (Kayseri Cevre Dostlari Dernegi) decided to start an education program and a documentary film on the benefits of the lake and threaths to the ecosystem with much success and increased sensitivity both in villagers, local, and national government authorities. It created an elevated willingness to conserve the lake both locally and nationally.

*4.6. Fuzzy cognitive map simulations*

Fuzzy cognitive map simulations were carried out for the local people and NGO&Government officials stakeholder groups' social cognitive maps separately and for the social cognitive map that included all the stakeholders. Results of computations were examined for 15 dependent variables which were ecosystem balance, biodiversity, livelihood, economic difficulties, agriculture, fruit and vegetable production, irrigation, animal husbandry, salt extraction, carpet weaving, tourism, birds, hunting, conservation of Tuzla Lake and drainage of Tuzla Lake. These variables were determined based on their importance for stakeholder groups, their centrality and their number of times mentioned.

In the first no management option, all dependent variables were set to 1 at the start and were allowed to change and settle to a final value freely. Final results higher than 0.5 were defined as high and values lower than 0.5 were defined as low. Results from the social cognitive map which includes all stakeholders reveal that ecosystem balance, biodiversity, livelihood, economic difficulties, agriculture, tourism, birds and conservation of Tuzla Lake are higher than 0.5 (Figure 3). However animal husbandry, fruit and vegetable production, carpet weaving and hunting is lower than 0.5. These results indicate that if there is no management and Tuzla Lake ecosystem continued as today, ecosystem balance, biodiversity and birds will increase and hunting will decrease. Livelihood will increase however economic difficulties will also become more. Agriculture, salt extraction and tourism will be major economic activities supporting livelihood but animal husbandry, fruit and vegetable production and carpet weaving will no longer be economically feasible.

In the results from the local people's social cognitive map, ecosystem balance, biodiversity and conservation of Tuzla Lake stabilized at 0.5, which indicates that they will remain the same if there is no management. Local people think that livelihood, agriculture, salt extraction, tourism and hunting and drainage of Tuzla Lake will increase and animal husbandry, carpet weaving and birds will decrease in the future. According to the social cognitive map of NGO&Government officials, ecosystem balance, biodiversity, livelihood, agriculture, animal husbandry, tourism, birds and conservation of Tuzla Lake will increase in the future. However, economic difficulties, fruit and vegetable production, irrigation, hunting and drainage of Tuzla Lake will decrease.

*4.7. Results of "What-if" scenarios*

Based on stakeholder views 65 "What-if" scenarios were run on the social cognitive map, which included all stakeholders. The effect of these scenarios was shown for 15 independent variables (Figure 4 shows the most important positive and negative simulation effects on ecosystem balance).

According to "What-if" scenarios run on the social cognitive map, which included all stakeholders, ecosystem balance and conservation of Tuzla Lake are very much related to the conservation status of the lake. Stakeholders reveal that if Tuzla Lake is declared as Wildlife Conservation Area, ecosystem balance will increase and conservation will be achieved more effectively. Conservation will result in less hunting. Another scenario which stakeholders focused on is education. Stakeholders think that when education is increased, ecosystem balance, biodiversity and success of conservation practices will be positively affected. According to stakeholders increase in education also causes decreases in hunting and in demands of local people about drainage of the lake. Stakeholders are aware that drainage of Tuzla Lake is harmful for ecosystem balance and biodiversity and they indicate that education through the Tuzla Lake Documentary Film will decrease these demands. Another scenario is increases in ecotourism activities such as nature photography, nature sports and bird watching. Stakeholders think that these activities will increase birds and success of conservation activities. For example, if local people benefited from tourism or community development programs, they were more likely to have positive attitudes toward conservation areas in Nepal (Mehta and Heinen, 2001).

Simulation results indicate that stakeholders are aware of agriculture and irrigation decreasing ecosystem balance and Tuzla Lake biodiversity. However, stakeholders depend on agriculture for their livelihood. They indicate that they have economic difficulties that may be eliminated if Tuzla Lake is drained and the area is used for agriculture. Also they think that pesticide and fertilizer use in agriculture will decrease economic difficulties through increased agricultural production. Education has an important role for their livelihood because it decreases economic difficulties and causes increase in agriculture, irrigation, carpet business and tourism. According to stakeholders irrigation has a big effect on agricultural productivity. They believe that agriculture will increase when irrigation increases, which can be achieved by providing water from the Kizilirmak or Bahcelik Reservoir. Irrigation also has an effect on development of animal husbandry. They think that agriculture will decrease when an alternative economic activity is developed. This activity could be salt extraction, carpet business or tourism. Stakeholders want to benefit from salt economically. They indicate that if a salt processing cooperative is established, it will contribute to their livelihood. However they also state that to increase salt extraction, the salt price should increase and Tuzla Lake watershed should be conserved. After the completion of this study local people with the help of an NGO has received a GEF Small Grants to establish an Ecologically, Sociologically and Economically Sustainable Salt Extraction Cooperative (SGP, 2004). Carpet business and tourism are very much related to agriculture. These economic activities are impacted negatively by agriculture and irrigation since local people tend to prefer agriculture to other activities. Ecotourism could provide for some income and incentive for local people to conserve Tuzla Lake, however this would have to be properly managed to avoid many pitfalls including lack of money being retained by local people (Valentine and Budowski, 1997).

## 5. Conclusion

Tuzla Lake is one of the important wetlands of Turkey that has not been seriously impacted by large-scale irrigation and drainage projects. The fuzzy cognitive mapping analysis indicated that there are many other threats to the ecosystem, including agricultural intensification, overgrazing, hunting and salt extraction and demands of local people for the drainage of the lake. Many of these threats, such as agricultural activities, overgrazing, and hunting, are common to wetlands in other Mediterranean countries (Papayannis and Salathe, 1999; Benessaiah, 1988).

*Agricultural Activities*: Agriculture is the main economic activity in the villages. Agricultural activities including grazing are concentrated on the lakeshores and around marshes. Cereals and beets are cultivated in the closed basin of the lake, especially on the southern and eastern side. The area under cultivation is being progressively extended in the direction of the lakeshores. On the western side, the cultivated belt has already reached the lake, even to the extent that water level rise in April floods the lower parts of some of these fields (Schekkerman and van Roomen, 1993). Agriculture was the most central variable and among the most mentioned variables in the cognitive maps drawn by the stakeholders.

*Villagers' demand for the drainage of the lake*: Since there is a perceived lack of land for cultivation, there is a continuing demand from the villagers for the drainage of the lake. As a result of this demand DSI (State Hydraulic Works - Devlet Su Isleri) prepared a draft plan to drain the lake to the Kizilirmak (Red River). But this plan was not put into operation since it was not feasible. The villagers' demand for drainage is still continuing. This threat was in the cognitive maps of the all the stakeholder groups but it was not perceived as a threat by the local people, but rather seen as a means to increase their livelihood.

*Bahcelik Reservoir Project*: In 1995 DSI started the construction of the Bahcelik reservoir on the Zamanti River, 55 km south east of Tuzla Lake. The reservoir will irrigate 37,000 ha of Kayseri, including 10,000 ha around Tuzla Lake. Land immediately around the lake is not included due to poor soil quality. DSI claims it will build an interceptor around the lake to channel polluted drainage water directly to Kizilirmak River. It is unclear how DSI will maintain the lake's water level once the interceptor is ready and the lake is cut-off from its natural water sources. Also, as many examples from elsewhere in Turkey have demonstrated, farmers tend to take water from drains and interceptors to irrigate marginal areas outside the irrigation scheme. At Tuzla Lake, this could lead to the permanent loss of valuable wetlands and natural salt steppe around the lake (Magnin and Yarar, 1997). The fuzzy cognitive mapping showed that NGO and government officials believe increasing irrigation will decrease ecosystem health, but local people think that irrigation will increase their livelihoods.

*Overgrazing*: Animal husbandry is the other important economic activity around the lake. In spring 1988, maximum numbers of some 2000 cattle and 2500 sheep were counted in the villages (Schekkerman and van Roomen, 1993). Both sheep and cattle grazing are quite intensive. Intensive grazing has some negative impacts such as the destruction of vegetation around the lake by trampling of soil and vegetation. Animal husbandry was very important in the fuzzy cognitive maps of all the stakeholder groups.

*Hunting*: Hunting is very widespread around the lake, almost every family has a gun. Monitoring work is the responsibility of local decision makers and the jandarmerie but they are not very proactive and hunting continues during the seasons when hunting is prohibited. The fuzzy cognitive maps of the stakeholders showed that local people, NGO and government officials all recognize that hunting decreases birds and wildlife.

*Salt Extraction*: Salt has been extracted from the lake since time immemorial. Previously, there were large salt production pans managed by the government but it stopped in 1968 since this activity was not found economically feasible. Later villagers continued to take salt from the lake for their daily usage and local trade. Salt is especially important for animal husbandry now and almost all the people having animals use it. However, currently local scale salt extraction does not seem to damage lake biodiversity. If large-scale salt extraction came to be an industrial activity in the future as it has been in the past, it could be damaging. Salt extraction was in the fuzzy cognitive maps of all the stakeholder groups as an ecosystem good ad establishing a sustainable extraction regime is underway through local initiative.

*No Ramsar Status and Limited Conservation Status*: Tuzla Lake ecosystem has not been designated as a Ramsar site yet and only construction has been zoned out in a narrow periphery of the lake since 1993. The absence of a strong conservation status puts the wetland ecosystem into danger. The limited conservation status was considered in the fuzzy cognitive maps of the NGO and government officials but not in the maps of the local people.

In this study we used a fuzzy cognitive mapping approach to obtain stakeholder views of and desires for Tuzla Lake. The fuzzy cognitive mapping approach provided an effective instrument to determine the views of different stakeholder groups that were villagers, local decisionmakers, government officials and non-government organization representatives. In particular the fuzzy cognitive mapping analysis showed that NGO&Government official maps are most similar to each other and different from local people. However, there were also similarities in the maps of the local people and NGO&Government officials, such as the focus of both groups on livelihood issues. Based on the stakeholders' views we developed a draft environmental management plan that may be used to develop a strategic environmental management plan for Tuzla Lake in the future.

From the analysis it was clear that Tuzla Lake is not only important for biodiversity but it also impacts local people's livelihood. So a management plan implemented to conserve Tuzla Lake ecosystem should also focus on socio-economic and cultural aspects of the ecosystem. The focus on socio-economic and cultural aspects can be achieved through a participatory ecosystem management approach. This conclusion is similar to other studies of wetlands in Mediterranean countries, which indicate that wetlands are an important source of livelihood for local people and therefore conservation and sustainable use must go hand-in-hand (Papayannis and Salathe, 1999; Benessaiah, 1988).

*Overall goal statement of management plan:* The results from the cognitive mapping were used to determine an overall goal statement for the environmental management plan that is acceptable for the stakeholders. According to the cognitive mapping the most important variables for both of the stakeholder groups were livelihood, and activities that enhanced livelihood such as agriculture, animal husbandry and salt extraction. In addition, the cognitive mapping results indicated ecotourism and carpet weaving could be developed as alternative economic activities. In their cognitive maps, NGO&Government officials focused on the conservation of Tuzla Lake and education of local people. Therefore, the overall goal statement of the draft environmental management plan is "Tuzla Lake ecosystem conserved and sustainable use of the local people provided". This statement covers both ecological and socio-economic goals and reflects the desires of all the stakeholder groups. As it was stated before, conservation activities for Tuzla Lake should also consider economic and social factors. By providing for the needs of local people and addressing their livelihood concerns, conservation projects can be more sustainable (Borrini-Feyerabend, 1997).

*Goals and Activities:* All of the goals and activities in the draft environmental management plan were developed based on the results of the fuzzy cognitive mapping, including what were the most important variables for stakeholders and the modelling of "What-if" scenarios.

One of the goals is to strengthen the conservation status of Tuzla Lake. This was in the fuzzy cognitive maps of the NGO&Government officials. Although Tuzla Lake meets the Ramsar standards, it has not been declared a Ramsar site yet. Currently Tuzla Lake is about to be declared a Wildlife Protection Area by the Turkish government. This status may lead to eventual designation of this ecosystem as a Ramsar site. Strengthening the conservation status of Tuzla Lake supported by strong cooperation and collaboration between stakeholders will achieve conservation more effectively since it will make the conservation activities more legitimate and strong. Stakeholder participation in decision-making and management will also increase stakeholders' awareness of the ecosystem, which is necessary since interviews and cognitive mapping results indicated that most of the villagers and some government officials were not fully aware of the importance of Tuzla Lake ecosystem. For biodiversity conservation, capacity of local organizations should be strengthened and they should be included in the implementation of management plans.

Another goal is to prevent threats towards Tuzla Lake ecosystem. The fuzzy cognitive maps of stakeholders, showed that they all perceive that hunting reduces wildlife and birds. Hunting can be prevented through local people's education and making monitoring work more proactive. Other threats, such as agriculture and animal husbandry (if overgrazing occurs), cannot be prevented completely in favor of conservation since local people's livelihoods highly depend on these activities. The solution could be increasing sustainability of traditional economic activities or supporting other sustainable economic activities. For example, to achieve sustainable agriculture, water saving technologies and organic agriculture can be considered. In addition, a grazing management plan should be developed and implemented.

The fuzzy cognitive mapping and interviews with local people showed that they are interested in alternative economic activities such as carpet weaving, ecotourism, and sustainable salt extraction. Carpet weaving was widespread in the past but it decreased as a result of agricultural development. Local people are enthusiastic to participate in this activity if marketing opportunities are developed. Tuzla Lake provides a big ecotourism potential especially for bird watching, nature sports and nature photography. Ecotourism potential can be used in favor of local people if government invested in this activity. Ecotourism, however would need to be carefully managed both not to degrade the ecosystem and to allow local people to retain the economic benefits. With regards to sustainable salt extraction, activities are already in operation with the cooperation of NGOs and local people.

The activities mentioned above form a basis for the strategic environmental management plan to be developed in the future. The draft management plan, consists of the major issues and activities, and the need for further micro-plans (e.g., sustainable salt extraction, grazing management). These micro-plans will have to answer and be guided by other challenges posed in the overall management plan and will need to answer problems in further detail. Applicability of the management plan depends on the continued inclusion of stakeholder views and the participatory processes, such as stakeholder meetings. If in these meetings and in the development of micro-plans a greater nuance and complexity in information is needed, it might be necessary to go back and look at the original maps, where 208 variables have been defined, run more simulations or conduct further targeted interviews. The results of cognitive mapping analysis can help facilitate future stakeholder meetings (Özesmi and Özesmi, 2003) and support the management practices leading to conservation of Tuzla Lake.

**Acknowledgements**

This research was supported by Erciyes University Research Funds and Turkish Scientific and Technical Research Council (TUBITAK - YDABAG-100Y112). We would like to thank local people and all other stakeholders from Tuzla Lake for their participation in this study. Special thanks are due to Stacy Özesmi for helping at almost each step and reviewing the manuscript. We also thank Can Ozan Tan and Anlı Ataöv for providing comments on the manuscript.

**References**

Bauer, V., 1975. Simulation, evaluation and conflict analysis in urban planning. In: Baldwin, M. M. (Ed.), Portraits of Complexity: Applications of Systems Methodologies to Societal Problems, Batelle Institute, Columbus, pp.179-192.

Benessaiah, N., 1988. Mediteranean Wetlands Socioeconomic Aspects. Ramsar Convention Bureau, Gland, Switzerland.

Berkes, F., Folke, C., 1998. Linking social and ecological systems for resilience and sustainability. In: Berkes, F., Folke, C.(Eds.), Linking Social and Ecological Systems: Management Practices and Social Mechanisms for Building Resilience, Cambridge University Press, Cambridge, pp.1-25.

Borrini-Feyerabend, G., 1997. Beyond Fences: Seeking Social Sustainability in Conservation (Vol. 1: A Process Companion). IUCN, Gland, Switzerland.

Bougon, M., Weick, K., Binkhorst, D., 1977. Cognition in organizations: an analysis of the Utrecht Jazz Orchestra. Administrative Science Quarterly 22, 609-639.

Buchy, M., Race, D., 2001. The twists and turns of community participation in natural resource management in Australia: what is missing?. Journal of Environmental Planning and Management 44, 293-308.

Colwell, R. K., 1997. EstimateS: Statistical estimation of species richness and shared species from samples. Version 5. User's Guide and application published at: http://viceroy.eeb.uconn.edu/estimates.

Dadaser, F., Özesmi, U., 2002. Stakeholder analysis for Sultan Marshes ecosystem: A fuzzy cognitive approach for conservation of ecosystems. EPMR2002, Environmental Problems of the Mediterranean Region, 12-15 April 2002, Nicosia, North Cyprus.

De Lopez, T. T., 2001. Stakeholder management for conservation projects: a case study of Ream National Park, Cambodia. Environmental Management 28, 47-60.

Eden, C., Ackerman, F., Cropper S., 1992. The analysis of cause maps. Journal of Management Studies 29, 309-323.

Filho, W.L., 1997. Integrating environmental education and environmental management. Environmental Management and Health 8, 133-135.

Finlayson, A. C., McCay, B. J., 1998. Crossing the threshold of ecosystem resilience: the commercial extinction of Northern Cod. In: Berkes, F., Folke C. (Eds.), Linking Social and Ecological Systems: Management Practices and Social Mechanisms for Building Resilience, Cambridge University Press, Cambridge, pp. 311-338.

Glicken, J., 1999. Effective public involvement in public decisions. Science Communication 20, 298-327.

Glicken, J., 2000. Getting stakeholder participation 'right': a discussion of participatory processes and possible pitfalls. Environmental Science and Policy 3, 305-310.

Hage, P., Harary, F., 1983. Structural Models in Anthropology. Oxford University Press, New York.

Harary, F., Norman, R. Z., Cartwright, D., 1965. Structural Models: An Introduction to the Theory of Directed Graphs. John Wiley&Sons, New York.

Hijortsø, C.N., 2004. Enhancing public participation in natural resource management using Soft OR – an application of strategic option development and analysis in tactical forest planning. European Journal of Operation Research 152, 667-683.

Hobbs, B.F., Ludsin, S.A., Knight, R.L., Ryan, P.A., Biberhofer J., Ciborowski, J.J.H., 2002. Fuzzy cognitive mapping as a tool to define management objectives for complex ecosystems. Ecological Applications 12, 1548-1565.

Hughey, K.F.D., Cullen, R., Kerr, G.N., 2000. Stakeholder groups in fisheries management. Marine Policy 24, 119-127.

Johnson, N., Ravnborg, H.M., Westermann, O., Probst, K., 2001. User participation in watershed management and research. Water Policy 3, 507-520.

Kapoor, I., 2001. Towards participatory environmental management? Journal of Environmental Management 63, 269-270.

Korfmacher, K.S., 2000. What's the point of partnering? American Behavioral Scientists 44, 548-564.

Korfmacher, K.S., 2001. The politics of participation in watershed modelling. Environmental Management 27, 161-176.

Kosko, B., 1987. Adaptive inference in fuzzy knowledge networks. Proceedings of the First IEEE International Conference on Neural Networks (ICNN-86), pp. 261-268, San Diego, California.

Kuniyal, J.C., Jain, A. P., Shannigrahi, A.S., 1998. Public involvement in solid waste management in Himalayan trails in and around the Valley of Flowers, India. Resources, Conservation and Recycling 24, 299-322.

Laszlo, E., Artigiani, R., Combs, A., Csanyi, V., 1996. Changing Visions, Human Cognitive Maps, Past, Present, and Future. Praeger, Westport, Connecticut.

Lise, W., 2000. Factors influencing people's participation in forest management in India. Ecological Economics 34, 379-392.

MacDonald, N., 1983. Trees and Networks in Biological Models. John Wiley&Sons, New York.

Magnin, G., Yarar, M., 1997. Important Bird Areas In Turkey. stanbul: DHKD.

Margerum, R.D., 1999. Integrated environmental management: the foundations for successful practice. Environmental Management 24, 151-166.

Makoloweka, S., Shurcliff, K., 1997. Coastal management in Tanga, Tanzania: a decentralized community-based approach. Ocean & Coastal Management 37, 349-357.

Mehta, J.N., Heinen, J.T., 2001. Does community-based conservation shape favorable attitudes among locals? An empirical study from Nepal. Environmental Management 28, 165-177.

Mendoza, G.A., Prahbu, R., 2003. Qualitative multi-criteria approaches to assessing indicators of sustainable forest resource management. Forest Ecology and Management 174, 329-343.

Ogu, V. I., 2000. Stakeholders' partnership approach to infrastructure provision and management in developing world cities: lessons from the Sustainable Ibadan Project. Habitat International 24, 517-533.

Özesmi, U., 2003. Dogal Alan Yonetimi: Sonsuz Ortaklik, Yesil Atlas 6, 64-69

Özesmi, U., 2001a. Bilissel (Kognitif) Haritalamaya Gore Halkin Talepleri (The Wants and Desires of the Local Population Based on Cognitive Mapping), In: Aygül, C. (Ed.), Yusufeli Baraji Yeniden Yerlesim Plani (Yusufeli Damlake Resettlement Plan), Devlet Su Isleri (DSI) (State Hydraulic Works), Sahara Muhendislik, Ankara, pp. 154-169.

Özesmi, U., 2001b. Uluabat Golu'nde Sulakalan-Insan Iliskileri (Uluabat Lake People-Wetland Relations and Stakeholder Group Analysis). DHKD Yayini, Ankara.

Özesmi, U., 1999a. Conservation Strategies for Sustainable Resource Use in the Kizilirmak Delta in Turkey, Ph.D. Dissertation, University of Minnesota, St Paul, 230 pp. http://env.erciyes.edu.tr/Kizilirmak/UODissertation.html

Özesmi, U., 1999b. Modeling ecosystems from local perspectives: fuzzy cognitive maps of the Kizilirmak Delta wetlands in Turkey. 1999 World Conference on Natural Resource Modelling, 23-25 June 1999, Halifax, Nova Scotia, Canada.

Özesmi, U., Özesmi, S. 2003. A participatory approach to ecosystem conservation: fuzzy cognitive maps and stakeholder group analysis in Uluabat Lake, Turkey. Environmental Management 31, 518-531.

Özesmi, U., Özesmi, S., 2004. Ecological models based on people's knowledge: a multi-step fuzzy cognitive mapping approach. Ecological Modelling 176, 43-64.

Palerm, Juan R., 1999. Public Participation in Environmental Impact Assessment in Hungary: Analysis through Three Case Studies, Environmental Impact Asssessment Review 19, 201-220.

Papayannis, Th., Salathé, T., 1999. Mediterranean Wetlands at the Dawn of the 21$^{st}$ Century. MEDWET, Tour du Valat.

Radomski, P. J., Goeman, T. J., 1996. Decision making and modeling in freshwater sport-fisheries management. Fisheries 21, 14-21.

Ribot, J. C., 1995. From exclusion to participation: turning Senegal's forestry policy around?. World Development 23, 1587-1599.

Richardson, T., Dusik J., Jindrova, P., 1998. Parallel public participation: an inertia in decision-making. Environmental Impact Assessment Review 18, 201-216.

Sample, A., 1993. A framework for public participation in natural resource decisionmaking. Journal of Forestry 91, 22-27.

Schekkerman, H., Van Roomen, M. W. J., 1993. Migration of Waterbirds through Central Anatolia, Spring 1988. WIWO Report No:32, Zeist.

Scott, J. C., 1998. Seeing Like a State: How Certain Schemes to Improve the Human Condition Have Failed. Yale University Press, New Haven.

Simon, H. A., 1996. The Sciences of the Artificial. The MIT Press, Cambridge.

SGP, 2004. Palas Golu'nde ekolojik tuz cikarimi projesi (Ecological salt extraction project in Palas Lake). GEF Kucuk Destek Programi Turkiye Bulteni 2(4), 2.

Somuncu, M., 1996. Palas Ovası'nda co rafya gözlemleri (Geographical observations in Palas Plain). Ankara Üniversitesi Türkiye Co rafyası Ara tırma ve Uygulama Merkezi Dergisi 5, 183-211.

Valentine, P., Budowski, G., 1997. Ecotourism. In: Borrini-Feyerabend, G. (Ed.), Beyond Fences: Seeking Social Sustainability in Conservation (Vol. 2: A Resource Book), IUCN, Gland, Switzerland, pp. 86-90.

Vallentyne, J.R., Beeton, A.M., 1988. The 'ecosystem' approach to managing human uses and abuses of natural resources in the Great Lakes Basin. Environmental Conservation 15, 58-62.

Webler, T., Tuler, S., Shockey, I., Stern, P. and Beattie R., 2003. Participation by local government officials in watershed management planning. Society and Natural Resources 16, 105-121.

Figure 1. The accumulation curve shows the number of new variables per interview versus the number of interviews for Tuzla Lake.

Figure 2. Condensed social cognitive map of all stakeholders (n=44). Only the strongest connections are shown. The thickness of the lines indicate the relative strengths of the connections. Solid lines show positive connections, dashed lines negative connections.

Figure 3. Steady state conditions of selected variables based on the neural network calculations for the social cognitive map (n=44). Values above 0.5 indicate that the variables will increase in the future if things continue as they are now, according to the stakeholders' perceptions, while values below 0.5 indicate that the variables will decrease.

Figure 4. Effect of "What-if" scenarios on ecosystem health for the social cognitive map (n=44). The change in values of selected variables from their steady state values are shown. Positive numbers indicate that the variables will increase for the given"What-if" scenario while negative values indicate the variables will decrease as compared to the steady state, according to stakeholders' perceptions.

Table 1. Number of stakeholders whose cognitive maps are used in this study for each group. Their community, gender, age, and occupation are given to characterize the participants.

|  | Number | Male | Female | Age | Occupation or Position |
|---|---|---|---|---|---|
| Local Decision Makers | Total 10 | | | | |
|   Karahidir | 1 | 1 | | 34 | Village Head |
|   Cavlak | 1 | 1 | | 40 | Village Head |
|   Ömerhacili | 1 | 1 | | 36 | Village Head |
|   Palas | 4 | 4 | | 44-55 | Village/Municipal Head |
|   Tuzhisar | 3 | 3 | | 45-54 | Village/Municipal Head |
| Villagers | Total 24 | | | | |
|   Karahidir | 5 | 3 | 2 | 32-76 | Farmer, Retired |
|   Cavlak | 2 | 2 | | 38-40 | Farmer, Retired |
|   Ömerhacili | 7 | 5 | 2 | 31-72 | Farmer, Retired |
|   Palas | 6 | 4 | 2 | 30-41 | Farmer, Retired |
|   Tuzhisar | 4 | 4 | | 45-80 | Farmer, Retired |
| NGO Representatives | Total 5 | | | | |
|   Bird Research Society | 1 | | 1 | 24 | NGO Officer |
|   Friends of the Environment | 4 | 3 | 1 | 24-33 | NGO Officers |
| Government Officials | Total 5 | | | | |
|   State Hydraulic Works | 1 | 1 | | 48 | Administrator/Engineer |
|   Directorate of Forestry | 1 | 1 | | 51 | Administrator/Engineer |
|   Directorate of Environment | 1 | 1 | | 35 | Engineer |
|   State Village Works | 1 | 1 | | 48 | Administrator/Engineer |
|   Provincial Bank | 1 | 1 | | 55 | Administrator/Engineer |

Table 2. The comparison of similarity and dissimilarity in the variables the maps contain between pairs of stakeholder groups.

| Stakeholder Groups | Phi | Yule Q |
|---|---|---|
| Local Decision Makers - Villagers | 0.417618* | 0.724315** |
| NGO – Government Officials | 0.264084* | 0.512064** |
| Local Decision Makers – NGO | -0.196531 | -0.381960 |
| Local Decision Makers – Devlet | -0.037248 | -0.078341 |
| Local Decision Makers - NGO& Government Officials | -0.257276 | -0.503033 |
| Villagers – NGO | -0.213425 | -0.420849 |
| Villagers - Government Officials | -0.022462 | -0.048458 |
| Villagers – NGO&Government Officials | -0.245470 | -0.500000 |
| Local People – NGO | -0.407404 | -0.780538 |
| Local People - Government Officials | -0.167587 | -0.369441 |
| Local People – NGO&Government Officials | -0.488373* | -1.000000** |

*The Phi Value indicates degree of similarity, where 1 is most similar.

**Yule Q Coefficient is the proportionate reduction in errors in predicting whether or not one group has the variable based on the knowledge that the other group has that variable.

Table 3. Graph theory indices calculated for each stakeholder group. *Statistically significant difference (Mann-Whitney test, p=0.0369). **Statistically significant difference (Mann-Whitney test, p=0.0143).

| | Local Decision Makers | Villagers | NGO | Government Officials | Local People | NGO&Government Officials |
|---|---|---|---|---|---|---|
| No. of Maps | 10 | 24 | 5 | 5 | 34 | 10 |
| | Mean±SD | Mean±SD | Mean±SD | Mean±SD | Mean±SD | Mean±SD |
| No. of Variables | 26.3±9.0 | 22.2±5.9 | 37.4±8.7 | 22.4±7.2 | 23.4±7.1 | 29.9±10.9 |
| No. of Receiver Variables | 8.9±4.4 | 7.5±2.7 | 8.0±1.9 | 6.4±2.7 | 7.9±3.3 | 7.2±2.3 |
| No. of Transmitter Variables | 10.9±3.8 | 8.0±2.4 | 12.6±6.8 | 9.6±2.9 | 8.9±3.1 | 11.1±5.2 |
| No. of Ordinary Variables | 6.5±2.9 | 6.8±3.2 | 16.8±9.1 | 6.4±2.8 | 6.7±3.1 | 11.6±8.4 |
| No. of Connections | 26.9±9.3 | 23.8±8.7 | 82.8±55.9 | 24.2±8.0 | 24.7±8.9* | 53.5±48.7* |
| Connection/Variable | 1.01±0.08 | 1.04±0.16 | 2.11±1.04 | 1.07±0.20 | 1.03±0.14** | 1.59±0.89** |
| Complexity Receiver/Transmitter | 0.849±0.479 | 1.025±0.533 | 0.849±0.522 | 0.655±0.239 | 0.973±0.517 | 0.752±0.396 |
| Density | 0.046±0.024 | 0.051±0.013 | 0.058±0.022 | 0.054±0.020 | 0.050±0.017 | 0.056±0.020 |
| Hierarchy | 0.035±0.047 | 0.037±0.028 | 0.047±0.051 | 0.024±0.018 | 0.036±0.034 | 0.036±0.038 |

Table 4. The most mentioned variables in the maps of stakeholders. Local people and NGO&Government officials share five of the ten most mentioned variables.

|    | Social (includes all stakeholders) | Local People | NGO&Government officials |
|----|------|------|------|
| 1  | **Livelihood** | **Livelihood** | **Livelihood** |
| 2  | *Agriculture* | *Agriculture* | *Agriculture* |
| 3  | **Animal Husbandry** | **Animal Husbandry** | **Animal Husbandry** |
| 4  | *Salt Extraction* | *Salt Extraction* | *Salt Extraction* |
| 5  | *Hunting* | Using lake salt in animal husbandry | Birds |
| 6  | Birds | Using lake salt in foods | *Hunting* |
| 7  | Using lake salt in animal husbandry | Using mud for rheumatism | Irrigation |
| 8  | Mud | Mud | People and Interactions with the lake |
| 9  | Tuzla Lake | *Hunting* | Education |
| 10 | Using lake salt in foods | Tuzla Lake | Conservation of Tuzla Lake |

Table 5. The most central variables in the social cognitive map of all stakeholders. High centrality indicates a variable that has great importance in the cognitive map. High indegree indicates that the variable is affected very much by other variables. High outdegree indicates that the variable affects other variables very much.

|  | Centrality | Indegree | Outdegree |
|---|---|---|---|
| Agriculture | 3.82 | 2.80 | 1.03 |
| Livelihood | 2.96 | 2.96 | 0.00 |
| Animal Husbandry | 2.70 | 2.05 | 0.65 |
| Tuzla Lake | 1.94 | 0.84 | 1.10 |
| Drought | 1.31 | 0.07 | 1.24 |
| Education | 1.22 | 0.13 | 1.09 |
| Salt Extraction | 1.14 | 0.64 | 0.50 |
| Hunting | 1.09 | 0.46 | 0.63 |
| Birds | 1.05 | 0.86 | 0.18 |
| Water level rise | 0.98 | 0.40 | 0.58 |
| Tourism | 0.88 | 0.69 | 0.18 |
| Irrigation | 0.85 | 0.36 | 0.48 |

Table 6. The ten most central (most important) variables in the social cognitive maps of the stakeholders. Local people and NGO&Government Officials share four of the ten most central variables in their social cognitive maps.

| No. | Local people | NGO&Government Officials |
|---|---|---|
| 1 | **Agriculture** | Education |
| 2 | *Animal Husbandry* | **Agriculture** |
| 3 | <u>**Livelihood**</u> | <u>**Livelihood**</u> |
| 4 | Tuzla Lake | Conservation of Tuzla Lake |
| 5 | Drought | Tuzla Lake Documentary Film |
| 6 | Salt Extraction | *Animal Husbandry* |
| 7 | Fog Formation | Tourism |
| 8 | Salt | Ecosystem Balance |
| 9 | Air Pollution | Hunting |
| 10 | Water Level Rise | Biodiversity |

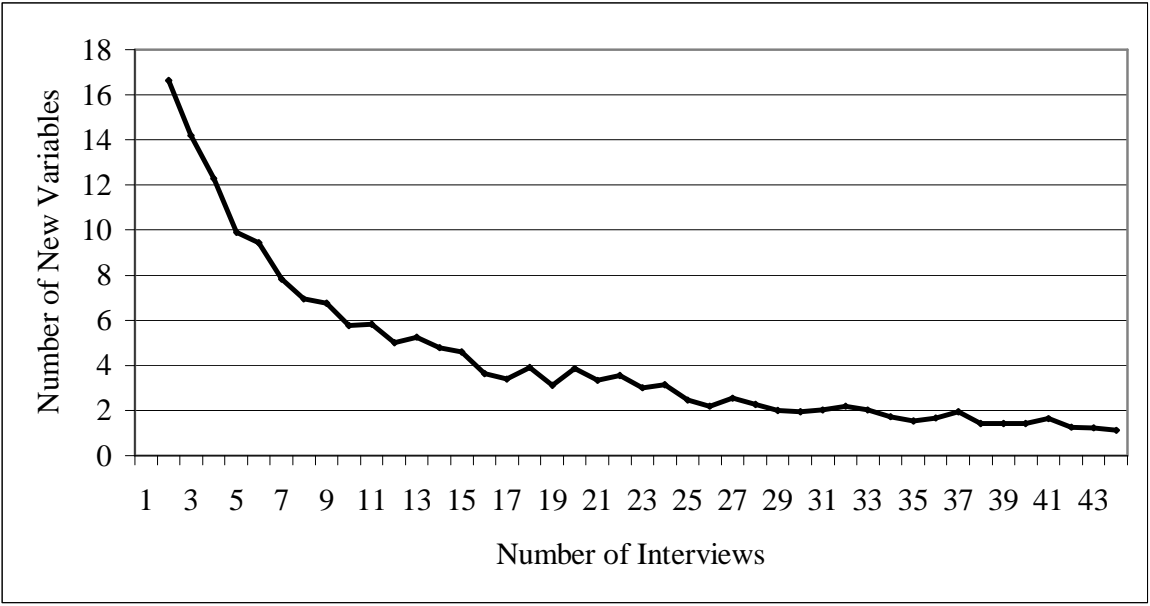

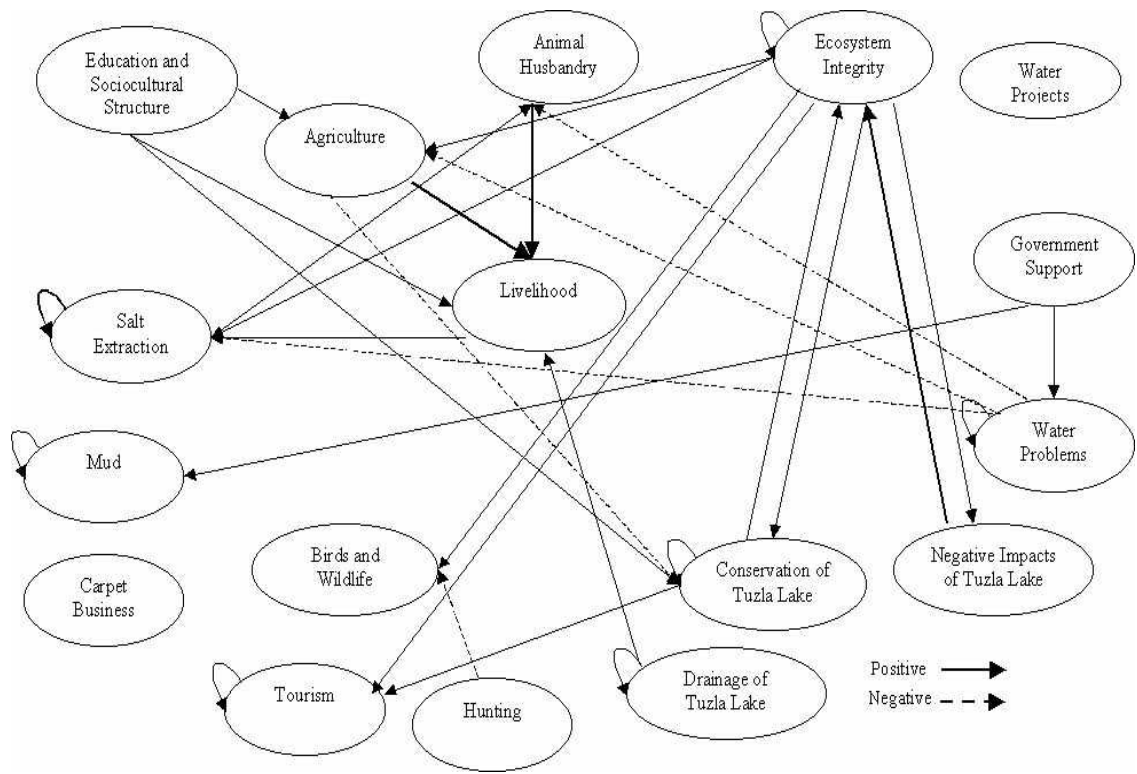

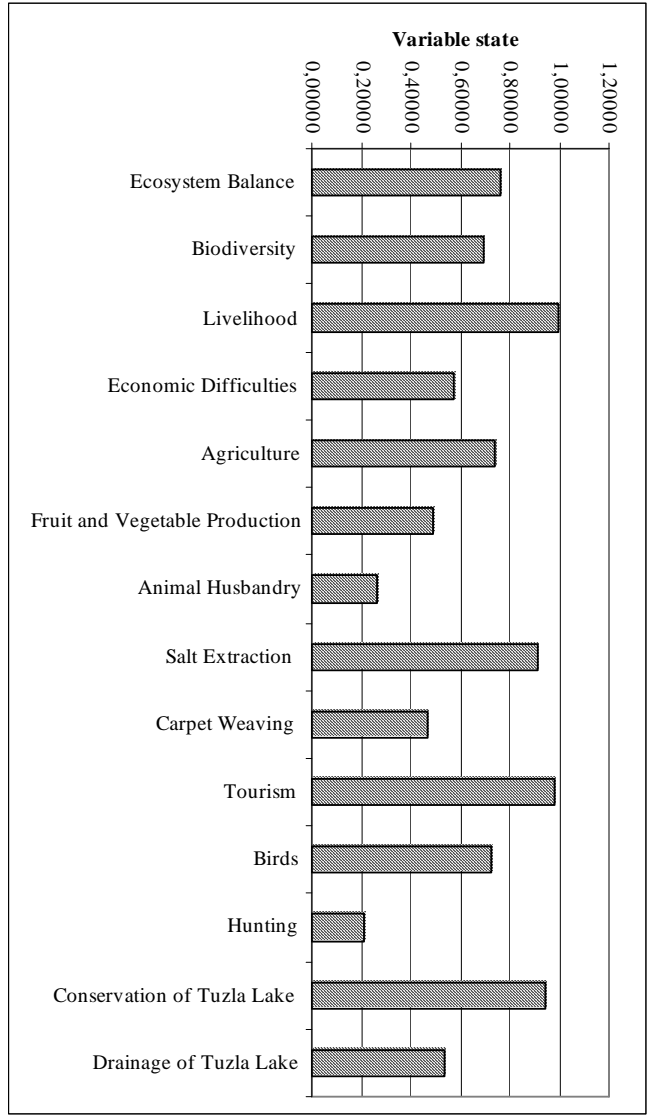

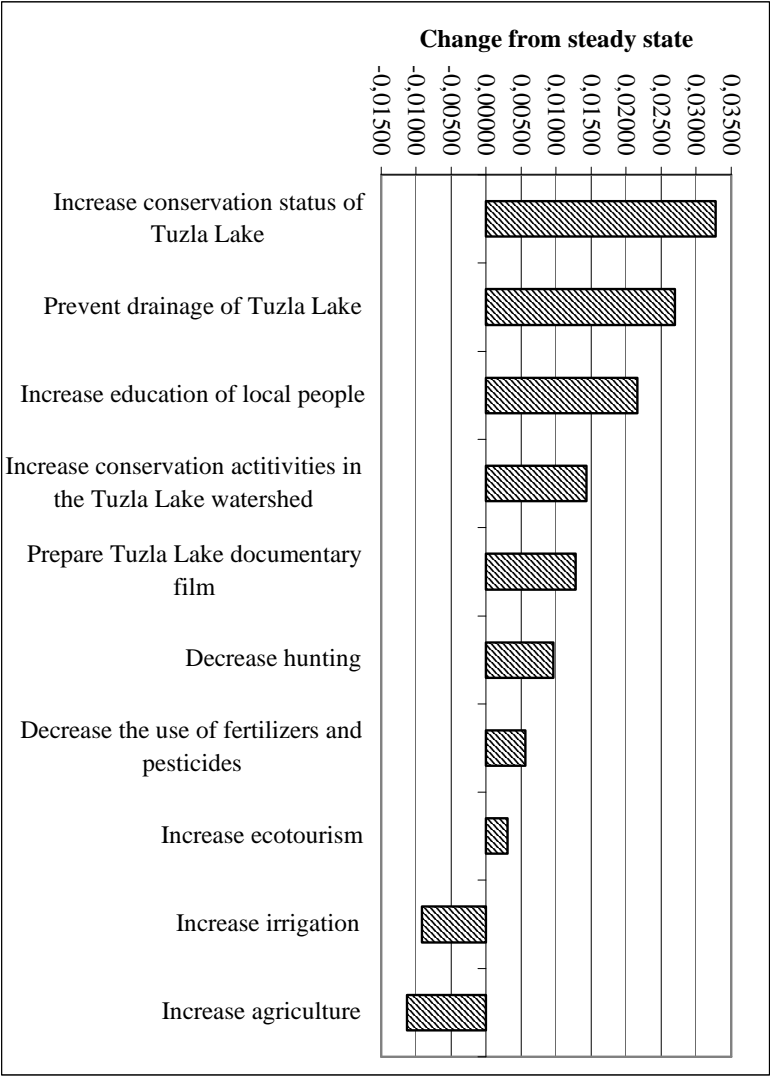